\documentclass{article}

\usepackage{authblk}
\usepackage{amssymb}
\usepackage{mathtools}
\usepackage{graphicx}
\usepackage[a4paper, portrait, margin=1.5cm]{geometry}
\usepackage{hyperref}
\usepackage{tikz}
\usepackage{braket}

\usepackage{microtype}

\title{Maritime Just-in-time navigation with Quantum algorithms}

\author[1]{Matthias Imrecke}
\author[2]{Fabian Klos}
\author[2]{Wolfgang Mergenthaler}
\author[2]{Michael Nowak}
\author[2]{Julian W\"uschner}

\affil[1]{Verband Deutscher Reeder, Ext. Advisor Nautical/Technical Affairs}
\affil[2]{Frankfurt Consulting Engineers GmbH, HOLM, \authorcr \textit{Bessie-Coleman-Str. 7, D-60549 Frankfurt am Main, Germany}}

\date{}

\begin{document}

\twocolumn[{%
  \begin{@twocolumnfalse}
    \maketitle
    \begin{abstract}
    	Just-in-time arrival in the maritime industry is a key concept for the reduction of Greenhouse gas emissions and cost-cutting, with the aim to reach the industrywide overall climate goals set by the International Maritime Organization (IMO) for 2030. In this note, we propose a mathematical formulation which allows for an implementation on quantum computers.
    \end{abstract}
    \vspace{.6cm}
  \end{@twocolumnfalse}
}]

\section{Just-in-time arrival}

Being a backbone of today's world logistics, the maritime industry's growth is accellerating and its organizations research technological innovation on all levels. One of those technological advances is Just-in-time arrival, which is an integral part of the industry's attempt to reach the climate targets set by the IMO for 2030.

\begin{figure*}
  	\centering
  	\includegraphics[width=180mm]{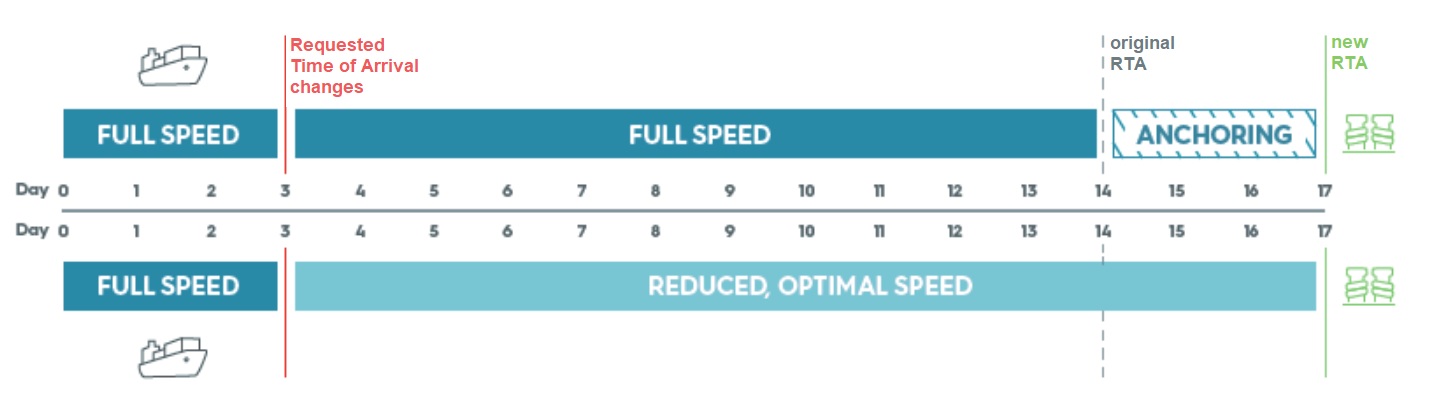}
  	\caption{An example for Just-in-time navigation. Above an example for ``Today's Operation: hurry up and wait''. Below an adapted version with ``Just-in-Time operation''. This image is taken from \cite[p. 3]{gia}.}
  	\label{fig:JIT-example}
  \end{figure*}

The basic idea is straightforward, and was for example discussed by the General Industry Alliance in \cite{gia}: Consider a vessel which is expected to reach its destination at the requested time of arrival (RTA) by driving full speed, see the upper row of figure \ref{fig:JIT-example}. At some stage during the trip, the destination authority changes the requested time of arrival to some later point in time. If the vessel's speed is kept at full speed, it arrives too early and needs to anchor, which causes extra costs and pollution at the destination. We have depicted an example of such anchoring costs in figure \ref{fig:harbor costs}.

A better solution is to reduce the speed as soon as the RTA changes - saving both fuel and anchoring costs. This is depicted in the second row of figure \ref{fig:JIT-example}.

This seemingly simple example contains a lot of analytics. To see this, note that the surrounding current's speed and direction along the path has to be taken into account. This is particularly interesting, because of its variation in time. On the one hand, the current's speed parallel to the vessel's path influences the estimated arrival time, but does \'a priori not have much bearing on fuel consumption. The perpendicular part, on the other hand, has to be counterbalanced and has a direct effect on fuel costs.

Overall, wind and current forecasts influence the decision which velocity should be adopted on which path segment. Every time the forecast or the requested time of arrival changes, the optimal velocities need to be adapted.

  \begin{figure}
  	\centering
  	\includegraphics[width=80mm]{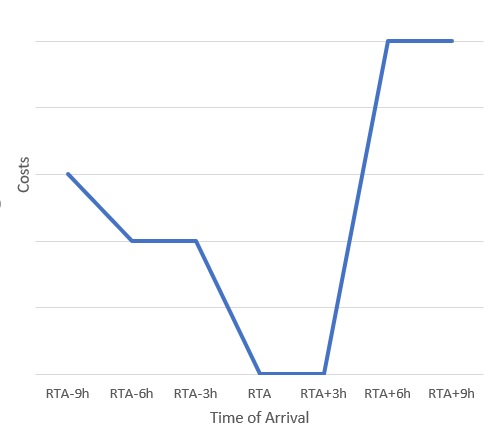}
  	\caption{Example for costs at destination as a function of ``time of arrival'' in relation to RTA.}
  \end{figure}
  	\label{fig:harbor costs}
  
  While the previous paragraphs only discussed speed variations, it is also possible to vary the vessel's path itself. 
  For a quantum version of such routing optimization problems with a view towards aircraft navigation, see \cite{jaroszewski2020ising}.
  
  For mathematical formulations of a different set of problems in maritime routing see \cite{9314905}.

\section{An Ising formulation}

Quantum adiabatic optimization \cite{farhi2000quantum, farhi2002quantum, crosson2014different}, the Quantum Approximate Optimization Algorithm \cite{farhi2014quantum} and related algorithms \cite{Peruzzo_2014, McClean_2016} on quantum hardware \cite{Preskill_2018, farhi2017quantum} allow to find the minimum of a quadratic unconstrained binary optimization (QUBO) polynomial
\begin{align*}
	C:\{0,1\}^n&\rightarrow\mathbb{R}_{\geq 0} \\
	\{X_i\}\;\,&\mapsto C(X_1, ..., X_n).
\end{align*}
For a collection of such polynomials see \cite{IsingOverview}. In the following, we develop such an optimization polynomial whose minimum encodes a solution to a simple JIT-optimization problem. At the end of this section, we point out ways to extend this formulation to more sophisticated real-world applications.

First, consider a fixed route of length $s$ divided into $n$ sectors $s_i$ with
\[
	\sum_{i=1}^n s_i = s.
\]
Let $\Delta v_i$ be the water flow rate parallel to the vessel route. Boost is indicated by $\Delta v_i > 0$. Perpendicular flow rates are commented on at the end of this section. Indicating the vessel's velocity relative to water in sector $i$ by $v_i$, the time for passing sector $i$ is given by
\[
	t_i = \frac{s_i}{v_i + \Delta v_i}.
\]
We now approximate the delay costs of figure \ref{fig:harbor costs} quadratically in the arrival time $t_A = t_1 + ... + t_n$. Denoting the fuel costs per time by a function $C_i$ quadratic in the velocities $v_i$, the total costs for given velocities $(v_1, ..., v_n)$ and requested time of arrival $\text{RTA}$ can be written down as
\begin{equation}\label{eq:poly}
	\sum_{i=1}^n C_i(v_i)\cdot t_i + \alpha (t_A - \text{RTA})^2.
\end{equation}
For the time being, we approximate $C_i$ linearly in $v_i+\Delta v_i$. The remaining second term is quadratic in the variables
\begin{equation}\label{eq:bin dec}
	(v_i + \Delta v_i)^{-1} := \sum_{j=-5}^0 2^j X_{i,j},
\end{equation}
which we have written in binary form.

\eqref{eq:poly} thus provides the quadratic polynomial whose minimum gives the optimal vector $(v_1, ..., v_n)$ solving the Just-in-time routing problem. This polynomial can now be fed to a quantum algorithm of choice.

In order to apply this method to real world applications, this model can be extended. Today's implementations of quantum optimization algorithms, e.g. by IBM or D-Wave, usually require the cost function to be maximally quadratic. However, \'a priori neither the quantum approximate optimization algorithm nor quantum adiabatic computing impose this constraint, see e.g. \cite{farhi2015quantum}. The costs per time $C_i$ in \eqref{eq:poly} can therefore be extended to be a function quadratic in the $(v_i+\Delta v_i)$ and additionally in the water velocities perpendicular to the vessels path.

Then, an additional penalty term relating the binary decomposition of \eqref{eq:bin dec} and its inverse is required:
\[
	P\cdot \left( (v_i+\Delta v_i) \cdot (v_i+\Delta v_i)^{-1} -1 \right)^2
\]

Dropping the requiredment of maximally quadratic terms also allows for more complex models. For example, instead of finding velocities on fixed route sectors $s_i$, one could consider the length of a fixed number of sectors as optimization variable.

\section*{Acknowledgements}

This work is supported by the German national initiative PlanQK.

\appendix

\section{Review of Quantum optimization methods}

The following review of quantum optimization solvers has appeared in \cite{jaroszewski2020ising} and is repeated here for exposition.

\subsection{Adiabatic Quantum Computing}

Quantum computation by adiabatic evolution as proposed in \cite{farhi2000quantum} is an optimization algorithm running on dedicated Quantum hardware. The basic idea is the following.

Let $C(X_1, ..., X_B)$ be a quadratic optimization function in $B$ binary variables $X_i\in\{0,1\}$. Construct the problem Hamiltonian $\mathcal{H}_P$
\[
	\mathcal{H}_P \ket{X_0}...\ket{X_B} = C(X_1, ..., X_B) \ket{X_0}...\ket{X_B}.
\]
Prepare your Quantum system in an easy-to-construct ground state $\ket 0$ of a simple initial Hamiltonian $\mathcal{H}_I$. Finally, let the system evolve in time from $t=0$ to $t=T$ along some monotonic curve $s(t)\in[0,1], s(0)=0, s(T)=1,$ according to the Hamiltonian
\[
	\mathcal{H}(t) := \mathcal{H}_I\cdot (1-s(t)) + \mathcal{H}_P \cdot s(t).
\]
After the evolution, the system is in the state $U(T)\ket 0$ where the time-evolution operator $U$ is the solution to the Schroedinger equation with respect to $\mathcal H (t)$.

If the energy gap between ground state and first excited state is greater than zero throughout the evolution and $T$ is chosen large enough, $U(T)\ket 0$ is the ground state of $\mathcal{H}_P$ at $t=T$ by the adiabatic theorem.

\subsection{QAOA}

The Quantum Approximate Optimization Algorithm \cite{farhi2014quantum} is a hybrid quantum-classical algorithm approximating the adiabatic evolution on gate-model Quantum hardware. Its concept can be summarized as follows.

Trotterizing the time-evolution operator $U$ of adiabatic Quantum computing into $p$ steps gives
\begin{align*}
	U(T) &\approx \prod_{k=1}^p e^{-\frac i \hbar \cdot \mathcal{H}(k\cdot\delta t)\cdot \delta t} \\
	&\approx e^{-i \beta_p \mathcal{H}_I} e^{-i \gamma_p \mathcal{H}_P} ... e^{-i \beta_1 \mathcal{H}_I} e^{-i \gamma_1 \mathcal{H}_P}.
\end{align*}
In the second step, we have linearized the exponential by suppressing higher order commutators in the Baker-Campbell-Hausdorff formula. The parameters $\beta_k$ and $\gamma_k$ depend on the form of $s(t)$.

The individual factors in the trotterized form of $U(T)$ can easily be implemented as gates on a universal quantum computer. For fixed $p$, $\beta_k$ and $\gamma_k$, QAOA evaluates the trotterized version of
\begin{equation}\label{eq:QAOA evaluation}
	\left(\bra 0  U(T)^\dagger \right) \hat C \left(U(T)\ket 0 \right)
\end{equation}
on a quantum computer. A classical optimization algorithm (e.g. gradient descent) now varies $\beta_k, \gamma_k$ while treating $p$ as a fixed hyperparameter. For every set of parameters the quantum computer evaluates \eqref{eq:QAOA evaluation} until the classical algorithm terminates. A Quantum measurement of $U(T)\ket 0$ for the final parameters $\beta_k, \gamma_k$ reveals the state minimizing the optimization function $C$.

\bibliographystyle{unsrt}
\bibliography{references}

\end{document}